\documentclass[twoside]{article}
\usepackage{fleqn,espcrc2}

\usepackage{graphicx}
\usepackage{epsfig}

\newcommand{\AmS}{{\protect\the\textfont2
  A\kern-.1667em\lower.5ex\hbox{M}\kern-.125emS}}

\newcommand {\bi} {\bibitem}
 \newcommand {\be} {\begin{equation}}
\newcommand {\bea} {\begin{eqnarray} \nonumber }
\newcommand {\ee} {\end{equation}}
\newcommand {\eea} {\end{eqnarray}}
 \newcommand {\eps} {\epsilon}
 \newcommand {\si} {\sigma}
\newcommand {\de} {\delta}

\newcommand {\ga} {\gamma}

 \newcommand {\al} {\alpha}

\newcommand {\ba} {\overline}
\newcommand {\lan} {\langle}
\newcommand {\ran} {\rangle}

\newcommand {\cP}  {{\cal P}}

\newcommand {\bc} {\begin{center}}
\newcommand {\ec} {\end{center}}
\newcommand {\bd}{\begin{displaymath}}
\newcommand {\ed}{\end{displaymath}}

\newcommand {\sign} {\mbox{sign}}
\newcommand {\for} {\ \ \ \mbox{for}\ \ }

\def \form#1 {eq. (\ref{#1}) }
\def \parziale#1#2  {{\partial {#1} \over \partial {#2}}}

\begin{document}
\title{Physics of glassy systems}

\author{Giorgio Parisi\address{Dipartimento di Fisica, Universit\`a {\it La  Sapienza}\\
and INFN Sezione di Roma I\\
Piazzale Aldo Moro 2, 00185 Roma (Italy)}
\thanks{E-mail: giorgio.parisi@roma1.infn.it}
}

\begin{abstract}
 In this talk I present some of the recent theoretical results that have been obtained 
 on glassy systems like spin glasses or structural glasses. The physical principles at the 
 basis of the theory are explained in a simple language (without using replicas) and the 
 results are compared with large scale numerical simulations. Finally we introduce the 
 generalized fluctuation dissipation relation that can be directly tested in experiments 
 with the present day technology.
\end{abstract}

\maketitle

\section {Introduction}

In this talk I will take the point of view that glassiness (roughly speaking the 
appearance, when  decreasing the temperature, of a very large equilibration time, much 
longer of those that can be observed on human scale) is related to metastability and to the 
presence of many equilibrium states.

Replica theory \cite{PARBOOK,MAPARIRUZU} is the most powerful tool to deal with systems with many 
equilibrium states.  Here I will present the main results of replica theory stressing the basic 
ideas, which have been recently recognized to be {\sl stochastic stability} and {\sl separability}.

In this talk I will first elaborate on the relations among glassiness and metastability.  Later on I 
will introduce some models of glassy systems (spin glasses, tilings and structural glasses).  I will 
then present the general theoretical interpretation of these phenomena, which can become more 
quantitative using the two principles of stochastic stability and separability and I will report on 
some numerical simulations which support these ideas.  In the next section I will highlight the 
connection among this approach and the off equilibrium fluctuation dissipation relations.  Finally I 
will present my conclusions.

\section{Glassiness and metastability}

It is not simple to give a precise definition of glassiness. In this talk I will take the 
point of view that glassiness corresponds to the presence of metastability in an open 
region of parameter space, a new and unusual phenomenon.

Let us recall the usual case in which we have metastability. We consider a system that 
undergoes  a first order phase 
transition when we change a parameter. The simplest example is the ferromagnetic Ising mode:
the control parameter is the magnetic field $h$. At low temperature the magnetization 
$m(h)$ is given by $m(h)=m_{s} \sign(h)$ for small $h$ ($m_{s}$ being the spontaneous 
magnetization): the magnetization changes discontinuously at $h=0$ in the low temperature phase 
where $m_{s} \ne 0$.

If we slowly change the magnetic field from positive to negative $h$, we enter in a 
metastable region where the magnetization is positive, and the magnetic field is negative.  
The system remains in this metastable state for a quite large time, given by $\tau(h) 
\propto \exp( A/|h|^{\alpha})$, where $\alpha=d-1$.  When the observation time is of order of 
$\tau(h)$ the system suddenly jumps into the stable state.  This phenomenon is quite 
common: generally speaking we always enter 
into a metastable state when we cross a first order phase transition by changing some parameters.

We can also define a linear response susceptibility which is given by
\be
\beta \chi_{LR}=\lim_{h\to 0^+}\sum_{i}\lan \si(i)\si(0)\ran^{c} .
\ee

If we start with the the state where $m>0$ at $h=0$ and we 
add a  {\em positive} magnetic field $h$ at time 0, the linear response susceptibility is  equal to
\be
\chi_{LR}= {\lim_{t \to \infty}} {\partial\over\ \partial h} m(t,h),
\ee
$ m(t,h)$ being the magnetization at time $t$.  The linear response susceptibility is not equal to 
the equilibrium susceptibility that in this case is infinite:

\be
\chi_{eq}= {\partial\over \partial h} {\lim_{t \to \infty}}  m(t,h){\biggr |}_{h=0} =\infty.
\ee

This is the usual stuff that is described in books.  We claim that in glassy system there is an 
open region in parameter space where, if we change the parameters of the system (e.g. the magnetic 
field $h$) by an amount $\Delta h$, we have that $\chi_{LR}\ne \chi_{eq}$.  In the case of spin 
glasses we expect that for $|h|<h_{c}(T)$ we stay in the glassy phase ($h_{c}(T)$ increases when we 
decrease the temperature and there is a temperature $T_{c}$ such that $h_{c}(T_{c})=0$).

In this region
\bea
\Delta m(t) =\chi_{LR} \Delta h & \for &  1<<t<<\tau(\Delta h),\\
\Delta m(t) =\chi_{eq} \Delta h & \for &  \tau(\Delta h) << t,
\eea
where in some cases one  finds numerically that $\tau(\Delta h)$ is has a power like behaviour
(e.g. $\tau(\Delta h) \propto |\Delta h|^{-4}$).

 It is convenient to define $\chi_{irr}$ as 
{$\chi_{eq}=\chi_{LR}+\chi_{irr}$}.  The glassy phase is thus characterized by a non zero value of 
$\chi_{irr}$.  If we observe the system for a time less that $\tau(\Delta h)$ the behaviour of the 
system at a given point of the parameter space depend on the previous story of the system and a 
strong hysteresis effects are present.

The aim of the the theoretical study of glasses is, at least from my point of view, to get a 
theoretical understanding of these effects  and to arrive to a qualitative and quantitative control
of these systems. Before presenting these efforts it is convenient to describe some models which 
have a glassy behavior.

\section{Glassy models}

Generally speaking glassy systems can be divided into two categories:
\begin{itemize}
    \item Systems with quenched disorder, e.g. spin glasses.  \item Systems without quenched 
    disorder, which are often translational invariant systems, e.g. tilings and glass forming liquids.
\end{itemize}

This distinction in two categories is important: although the two categories behave in the same way 
in the mean field approximation, it is quite possible that in finite dimension they have some 
different features \cite{DIFF}. We will see later that the definition of an ensemble will be slightly different 
in the two classes of models.

\subsection{Spin glasses}

The simplest model of spin glasses \cite{PYBOOK} is described by an Hamiltonian
\be
H_{U}(\si)=-\sum_{i,k}U(i,k)\si(i) \si(k) -\sum_{i}h \si(i),
\ee
where $U(i,k)$ is a quenched $Z_{2}$ gauge field on the lattice at infinite temperature 
(in plain words the variables $U$ are defined on the links, take the values $\pm 1$ and 
are uncorrelated quenched random variables),
$\si({i})$ is a matter field which also belongs to $Z_{2}$.  We are interested to compute the average 
value of the free energy (and of the correlation functions) at a temperature $\beta^{-1}$ of the matter 
field:
\bea
F(\beta)=-\beta^{-1} \ba{\ln(Z_{U}(\beta)) }\\
Z_{U}(\beta)=\sum_{\{\si\}} \exp(-\beta H_{U}(\si)).
\eea
A few comments are in order:
\begin{itemize}
    
    \item The overbar denote the average over the $U$: we are in the quenched approximation.  The 
    unquenched case in not physically interesting.  
    
    \item The quantity $h$ is the magnetic field, which breaks gauge 
    invariance.  
    
    \item At $h=0$ ``to find the minimum of $H_{U}(\si)$'' is equal ''to find the Landau 
    gauge'', $\si$ being the gauge fixing. For given $U$ in more than two dimension the problem of 
    finding  the global minimum of $H_{U}(\si)$ is an NP complete problem.
    
    \item Gribov ambiguity (which follows from the NP completeness of the problem) implies that 
    there are many local minima of $H_{U}(\si)$ (i.e. configurations whose energy does not decrease when we flip 
    one spin); their number increase exponentially with the volume.
\end{itemize}

As we have already remarked there is a 
a glassy region for not too large magnetic fields at low temperature.

\subsection{Wang Tilings}

The definition of Wang tilings is quite simple \cite{WANG}.  In each point of the lattice 
there are variables $\si(i)$ which can take $M$ different values and are called tiles.  
The Hamiltonian is given by
\be
H(\si)=\sum_{i}\sum_{\mu=1,d} E_{\mu}(\si(i), \si(i+\mu)).
\ee

The model depend on the functions $E_{\mu}(\si, \tau)$, which we assume can take only 
the values 0 and 1.  For each $M$ the number of different models is $2^{DM^{2}}$, however some of 
them are  related by 
symmetries. Particular examples of Wang tilings are the Ising model ($M=2$), the $p$-state
Potts models ($M=p$), the Baxter vertex models.

If there are configurations such that $H(\si)=0$, we say that the model admits a perfectly matched 
tiling.

In two dimensions for sufficiently large  $M$ weird things may happen:
\begin{itemize}
    
    \item For $M\ge 16$ there are functions $E_{\mu}(\si, \tau)$ such that for open boundary there 
    are global minima of $H$ with $H=0$, but they cannot be periodic.  In a finite volume system 
    with periodic boundary conditions  there are no perfectly matched tilings.
    
    \item For $M\ge 56$ there are functions $E_{\mu}(\si, \tau)$ such that the problem of deciding if a 
    configuration of tiles on a finite volume can be imbedded in an infinite volume ground state 
    with $H=0$ 
    is not decidable.  In other words there is no computer program which can tell us in a {\sl a priori} 
    bounded time if a given configuration can be imbedded in an infinite volume perfectly matched tiling.  We can 
    trivially construct a computer program that stops only if the configuration cannot be imbedded in an 
    infinite volume ground state, however if $L$ is the size of the configuration, the maximum  time 
    before stopping increases faster than any computable function of $L$.
\end{itemize}

Not so much is known on the thermodynamical properties of those models \cite{LEPA}: most 
of the investigations deal with the structure of the ground state configurations.
\begin{figure}
    \epsfig{figure=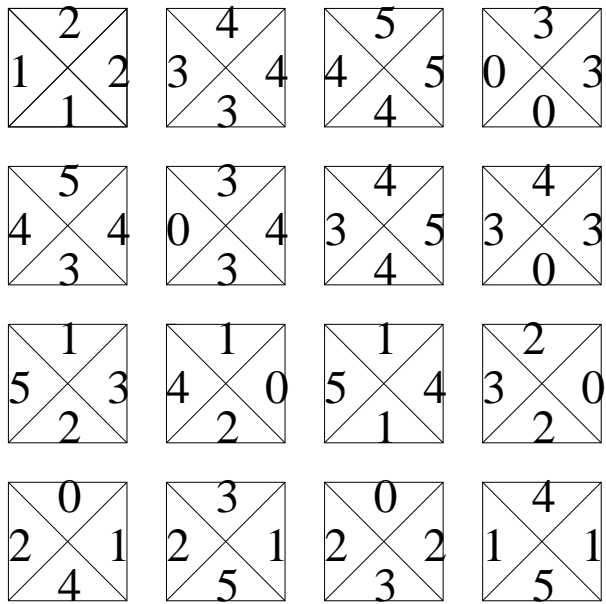,width=8cm}
    \caption{The 16 Wang tiles which lead to aperiodic tiling: $E_{\mu}(\si, \tau)$ is zero only if 
    the two faces in contact of two nearby tiles are equal.}
\end{figure}

\subsection{Glasses}

Simple models of glasses are particles of different $M$ species  with Hamiltonian
\be
H=\sum_{a,b=1,M}\sum_{i=1,N(a)}\sum_{k=1,N(b)} V_{a,b}(x_{a}(i)-x_{b}(k)),
\ee
where $N(a)=N c(a)$, $\sum_{a} c(a)=1$ and the quantities $c$ are the concentrations of each 
different species.  The model depends on the $M(M+1)/2$ functions $V_{a,b}(x)$ and on 
the $M$
concentrations.

In order to have a glass transition it is crucial that the model does not crystallize, i.e. that 
the crystal ground state energy is bigger than that of lowest energy amorphous structure.

A well studied model of glasses is realized for $M=2$, $c(1)=.8$, $c(2)=.2$ with a potential that 
has a simple Lennard-Jones form \cite{KB}.

\section{The theoretical interpretation}

One way to interprete the presence of metastable states is to assume that for large (but finite) systems the 
phase space of equilibrium configurations can be decomposed into many finite volume {\it states} or 
{\it valleys}
({\it lumps}, as suggested Talagrand \cite{TALA}).

In other word we have a breaking of the ergodicity: the Gibbs measure can be approximately 
written the sum of smaller disconnect pieces.  We warn the reader that we have to navigate 
between Scilla and Cariddi \cite{MAPARIRUZU}.  We cannot directly take the infinite volume 
limit because it is quite likely that the correlation functions do not have any infinite 
volume limit (at least in a naive sense). On the other hand, if we work on a very large, 
but finite, volume system, the notion of equilibrium state is physically intuitive, but it 
need a more sophisticated mathematic definition than the  usual notion of 
infinite volume pure states (such a definition can be given and it can be proved that the 
picture I am presenting rigorously holds, at least in some long range models \cite{TALA}).

Finite volume equilibrium states are  characterized by the following properties:

\begin{itemize}
    \item Connected correlation functions in each state go to zero al large distances ({\it cum grano 
    salis}).
    
    \item The time to go from one state to an other state is exponentially large with the volume.
\end{itemize}
    
In glassy systems one finds that these states at low temperature have the following unusual features
\begin{itemize}
\item Chaos: when we change the parameters in the Hamiltonian of any finite amount (typically as 
    soon as we make a change greater than $V^{-1/2}$, $V$ being the volume of the system)) we have 
    state crossing: stable states become metastable.
    
    \item No Gibbs rule: states coexist in an open region of parameter space.
\end{itemize}

The crucial step is  the decomposition  of the Gibbs measure for a finite system into states:
\be
\lan \cdot \ran=w_{\al} \lan \cdot \ran_{\al}\ ,
\ee
where the weights $w$ satisfy the relation $\sum_{\al}w_{\al}=1$.  Of course the previous formula 
is only approximate for a finite systems (there is a fraction of phase space with a small, but 
finite probability that cannot be classified into states) and should become more and more 
exact when the volume goes to infinity.

It is usually assumed that all the states have similar intensive properties, for example the 
internal energy, the
magnetization, the linear response susceptibility do not depend on the state:
\bea
N^{-1} \lan M\ran_{\al} = m+ N^{-1/2} \delta_{\al},\\
\chi_{\al}\equiv N^{-1}\lan M^{2}\ran ^{c}_{\al}=\chi_{LR}\ .
\eea
In this case we can also define the so called Edward Anderson parameter $q_{EA}=N^{-1}\sum_{i}
(\lan \si(i) \ran_{\al})^{2}$, which should not depend on the state $\al$.

However if we compute the full expression for the susceptibility, we find that the $O(N^{-1/2})$ 
variations in the magnetization do matter: the expression for $\chi_{irr}$ turns out to be:
\be\chi_{irr}=\sum_{\al,\ga} w_{\al}w_{\ga}(\de_{\al}-\de_{\ga})^{2} .\ee
\begin{figure}\label{UNO}
    \epsfig{figure=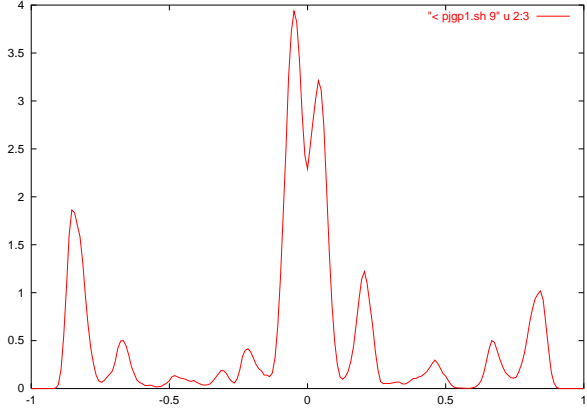,angle=270,width=8cm}
    \caption{The function $P_{J}(q)$ for one given $16^{3}$ sample.}
\end{figure}
The physical origine of this extra term is quite clear: when we increase the magnetic field, the states 
with higher magnetization become more likely than the states with lower magnetization and this 
effect  contributes to the increase in the magnetization. However the time to jump to a state to 
an other state is very high (it is strictly infinite in the infinite volume limit if non linear 
effects are neglected) and this effect produces the separation of time scales relevant for $\chi_{LR}$ and 
$\chi_{eq}$.

Having understood that the coexistence of many states in an open region of parameter space with the 
aforementioned properties is a crucial feature of glassy systems, we face the problem of being more 
quantitative. This quill be done in the next section.

\section{A quantitative approach}

\begin{figure}\label{DUE}
    \epsfig{figure=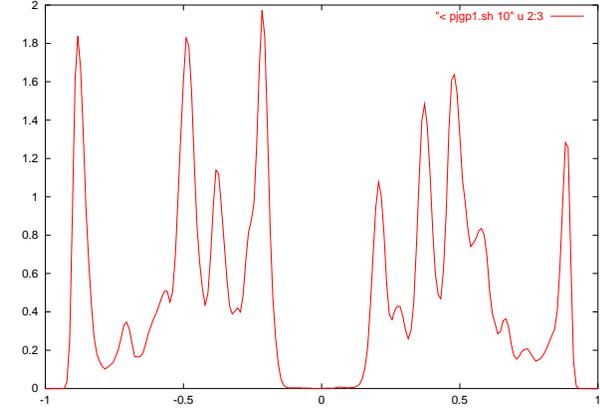,angle=270,width=8cm}
    \caption{The function $P_{J}(q)$ for an other $16^{3}$ sample.}
\end{figure}
In order to be specific, let us consider the case of spin glasses.  A very important object is the 
overlap among two configurations (e.g. $\si$ and $\tau$):
\be
q(\si,\tau)=N^{-1}\sum_{i}\si(i)\tau(i).
\ee
In a similar way we can define the overlap $q_{\al,\ga}$ among two states $\al$ and $\ga$ as the 
overlap among two generic configurations belonging to the two states. We also have:
\be
q_{\al,\ga} =N^{-1}\sum_{i}\lan \si(i) \ran_{\al}\lan \si(i) \ran_{\ga}.
\ee

In a similar way we can define a generalized overlap: given a local operator
$A(i)$ we have:
\be
q^{A}_{\al,\ga} =N^{-1}\sum_{i}\lan A(i) \ran_{\al}\lan A(i) \ran_{\ga}\ .
\ee
If we  take
$A(i)=\si(i)$ we get $q^{A}= q$ (i.e. the usual overlap):
if we take 
$A(i)=H(i)$ we get $q^{A}= q^{E}$ (i.e. the energy overlap).

Let us consider a given finite system characterized by some parameters that we call $J$.  In order 
to describe the structure of its states, we should give all the $w_{\al}$ and all the 
$q^{A}_{\al,\ga}$.  Equivalently we could introduce the probability distribution of the overlap  given by
\bea
P_{J}(q)=\\
Z^{-2}
\sum_{\{\si\},\{\tau\}} \exp(-\beta (H(\si)+H(\tau))\delta(q(\si,\tau) -q)\\
\approx
\sum_{\al,\ga} w_{\al}  w_{\ga} \delta (q-q_{\al,\ga}). \nonumber
\eea

\begin{figure}
    \epsfig{figure=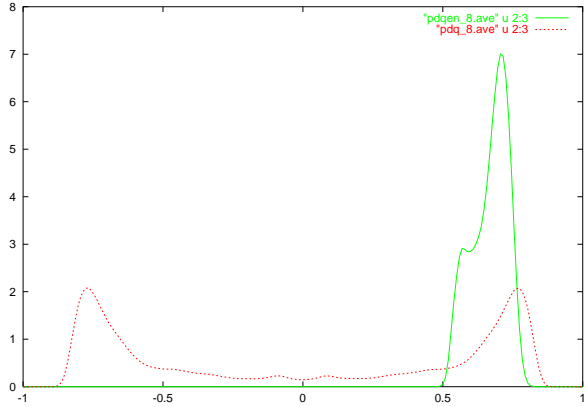,angle=270,width=8cm}
    \caption{The average value of the overlap distribution (dashed line) and the energy overlap 
    distribution (full line).}
\end{figure}

The structure of the states depends on the sample: we show in figs.  (\ref{UNO})-(\ref{DUE}) two 
$P_{J}(q)$ for two different samples of three dimensional spin glasses of size $16^{3}$.

Is is clear that this description (i.e. the function $P_{J}(q)$) depends on the sample and 
numerically the dependence does not disappear by increasing the size. In this case the only 
thing that makes sense is to introduce an ensemble and to describe the statistical properties of 
that ensemble.  At this end there are a few possibilities:
\begin{itemize}
    \item To average over the quenched disorder at fixed number of spins $N$.
    \item To average over $N$ in a window.
    \item To average over small quenched added disorder for a given sample.
\end{itemize}
It is evident that the first possibility is empty is no quenched disorder is present.

The statistical description would amount to assign the probability  $\cP (\{w\},\{q\})$ of finding a 
given configuration of weights $w$ and overlaps $q$. Such a functional is not so easily parametrized, 
so that we can introduce its moments, e.g.
\be
P(q) =\ba{P_{J}(q)},\ \ \ 
P(q_{1},q_{2}) =\ba{P_{J}(q_{1})P_{J}(q_{2)}}.
\ee

It is evident that $\cP (\{w\},\{q\}$) contains the same information of the  infinite set of functions 
($P(q),P(q_{1},q_{2}),P(q_{1},q_{2},q_{3})$.

In order to put order into this mess and to restrict the form of all possible $\cP (\{w\},\{q\}$
two principle can be assumed: {\sl stochastic stability} and {\sl separability}.

\begin{figure}\label{aaa}
    \epsfig{figure=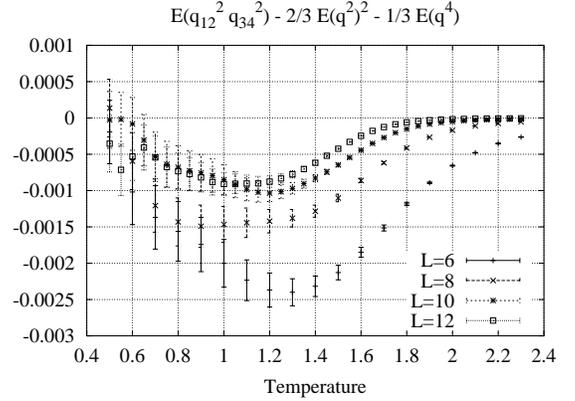,width=8cm}
    \caption{The quantity $\ba{q_{12}^{2}q_{34}^{2}} -2/3 \left(\ba{q_{12}^{2}}\right)^{2}-1/3 
    \ \ba{q_{12}^{4}}$.}
\end{figure}

\subsection{Stochastic stability}

Roughly speaking stochastic  stability  implies that the ensemble we consider 
is a generic random ensemble and it does not have any peculiar property \cite{G,AC,P}.  More technically 
it implies that if we add a random perturbation, i.e. we consider the Hamiltonian
\be
H(\eps)(\si)=H(\si)+\eps H_{R}(\si),
\ee
where $ H_{R}(\si)$ is a random perturbation, the expectation value of everything is smooth in 
$\eps$.  A typical example of random Hamiltonian we can add is 
$H_{R}(\si)=N^{-1/2}\sum_{i,k}R_{i,k}\si(i)\si(k)$, where $N$ is the total number of spin (as usual) 
and $R_{i,k}$ are random quantities with zero average and variance 1.  It is crucial in the argument 
that infinite range Hamiltonians are allowed as perturbation.
Stochastic stability implies very strong constraints on $\cP (\{w\},\{q\})$: the simplest one can be 
written under the form of the so called Guerra relations:
\be
P(q_{1},q_{2})= \frac23 P(q_{1})\delta( q_{1}-q_{2}) +\frac13 P(q_{1})P(q_{2}).
\ee
These relations are very well satisfied in short range model os spin glass. Indeed they imply
that
\be\ba{q_{12}^{2}q_{34}^{2}}=\frac23 \left(\ba{q_{12}^{2}}\right)  ^{2}-\frac13 \ba{q_{12}^{4}}.
\ee
This relation which is well satisfied numerically (see fig. (\ref{aaa}, where 
$\ba{q_{12}^{2}q_{34}^{2}}$ is a quantity of order one for $T < .9$).
As we shall see in the next section stochastic stability has important consequences on the dynamics.

Although for the moment we cannot prove that spin glasses and other random systems are 
stochastically stable, stochastic stability it is rather likely to be correct for system 
which do not have any symmetry (or if symmetries are present, it should hold if restricted to 
those observables which are left invariant by the action of the symmetry group).

\subsection{Separability}

Separability is a more subtle property \cite{P,PARI}.  It implies that there is only one 
kind of significant overlap.  For example in spin glasses it amounts to say all overlaps 
(depending on the operator $A$) are given functions of the spin overlap.  In other words 
for each local operator $A$ there is a function $f^{A}(q)$ (which may depends on the 
magnetic field, temperature \ldots) such that $q^{A}_{\al,\ga}=f^{A}(q_{\al,\ga)})$.

In the general case an infinite number of quantities (i.e. $q^{A}_{\al,\ga)}$ $\forall A$)
characterizes the mutual relations among $\al$ and $\ga$.
If separability is assumed this infinite number is reduced to one.

There are some arguments \cite{PARI} that suggest that separability implies the 
ultrametricity property:
\be
q_{\al,\ga)} > \min(q_{\al,\beta)},q_{\beta,\ga}) \ \forall \beta.
\ee
If ultrametricity holds, the set of states of a given system may be ordered on a tree (as 
it is usual in taxonomy) and the form of the probability functional $\cP (\{w\},\{q\})$ is 
very similar to the one found in the mean field approximation.

Separability and ultrametricity may be less compulsory properties than stochastic stability: this 
point is at present under investigations, however the validity of ultrametricity seems to be 
supported by numerical simulations for spin glasses, at least in 4 dimensions \cite{CAC}.

\section{Off-equilibrium dynamics}

A very interesting question is what happens when we cool the system starting from a random 
(high temperature) configuration at time zero?  The question is non trivial if, cooling 
the system, we cross a phase transition.  The behaviour of the system in these conditions 
tell us something about the order parameter in the low temperature phase (in the 
following we will discuss only the behaviour of an infinite volume system). One of the 
most spectacular phenomenon is ageing.  

Let us suppose that we cool very fast the system 
from an high temperature configuration at time 0.  We can define a two times correlation 
function
\bea
C(t,t_w) \\
\equiv \frac1N \sum_{1}^{N}  \hspace{1pt} _{ i}
\lan \si_i(t_w) \si_i(t_w+t)\ran =q(t_{w},t_{w}+t)
\eea

\begin{figure}\label{CORR}
    \epsfig{figure=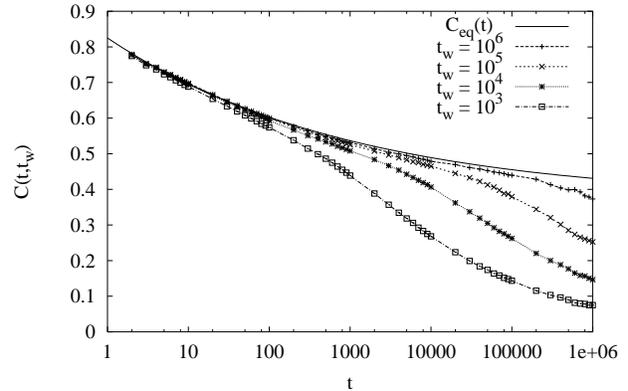,width=8cm}
    \caption{The correlation function $C(t,t_w)$ in 3-d spin glasses in the low temperature phase,}
\end{figure}

In the limit $t_{w}\to \infty$ at fixed $t$ we recover the usual equilibrium correlation 
function (i.e. $\lim_{t_{w}\to \infty}C(t,t_w)= C (t)$\ ).  However in the region 
where $t_{w}$ and $t$ are both large, the dependance on $t_{w}$ does not disappear: for 
example in fig.  \ref{CORR} we show the dependance of the correlation function on $t$ in 
four dimensional spin glasses: as soon as $t$ is of order $t_{w}$, the value of $t_{w}$ 
matters.  If simple aging holds, we have that $C(t,t_w) \approx {\cal C}(t/t_w)$ in this region.

We notice that $\lim_{t\to\infty}\lim_{t_w\to\infty}C(t,t_w)$ is a non zero quantity, 
(i.e. it is equal to $q_{EA}$) while $\lim_{t_w\to\infty}\lim_{t\to\infty}C(t,t_w)$ is 
zero at zero external magnetic field.  The non commutativity of the two limits is a very 
clear signal of existence of more than one equilibrium state.

\begin{figure}
    \epsfig{figure=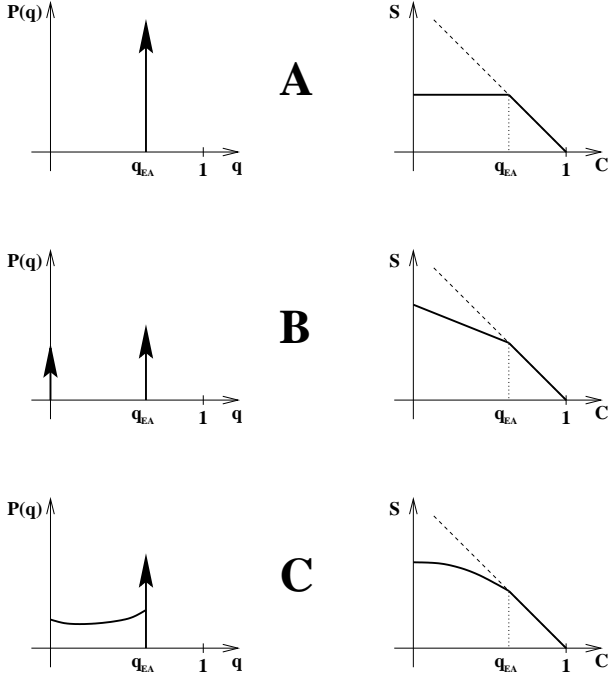,width=8cm}\label{TRE}
    \caption{Three different form of the function $P(q)$ and the related function $S(q)$}
\end{figure}

Aging can be intuitively interpreted in the following way.  At large times the system is ordered in 
one given phase in regions of size $\xi(t)$, where $\xi(t)$ goes to infinity when $t$ goes to 
infinity. These regions move with time and their coalescence produces the increase in the 
correlations length.  It is natural to suppose that 
\be
C(t,t_w)= f(\xi(t+t_w)/ \xi(t_w))
\ee
(a similar formula was suggested to be valid also in infinite range systems where no 
correlation length can be defined \cite{CUKU}).  The increase of the the dynamical 
correlation length $\xi(t)$ corresponds to rearrangements of the regions which belong to 
different phases (some shrink and some others expand) and it affects the two-times
correlation.  If $\xi(t)$ increases as power of time, the previous arguments suggest the 
validity of simple ageing.  Of course when $\xi(t)$ hits the size of the box ageing ends, 
however this phenomenon never happens for an infinity volume system if we use a local 
dynamics.

In a similar way we can consider a time dependent Hamiltonian
\be
H=H_{0}+\theta(t-t_{w}) \sum_{i} h_{i}\si_{i}\ ,
\ee
$h_{i}$ being a random magnetic field (which in spin glasses is gauge equivalent to a 
magnetic field with constant sign).

After cooling the system at time zero, we switch on the random field at time $t_{w}$. In this way we 
can define  a two times response function (the precise name is {\sl relaxation function}):
\be
\beta S(t,t_w) \equiv \frac1N \sum_{1}^{N}  \hspace{1pt} _{ i}
\lan {\partial \si_i(t_w+t)\over \partial h_{i}} \ran
\ee

A great simplification comes out if we close our eyes on the actual time dependence of the 
correlation function and of the relaxation function and we plot $S$ versus $C$ \cite{CUKU,CUKU1}.  
We can expect that such a plot has a well defined limit when $t_{w}$ goes to infinity.  In this
plot we must distinguish two regions:
\begin{itemize}
    \item The region $C>q_{EA}$ is the equilibrium region $t<<t_{w}$.  Here the fluctuation 
    dissipation theorem implies that
    \be
    -{dS \over dC} =1\ .
    \ee
    \item 
    The region $C<q_{EA}$ is the aging region where the fluctuation dissipation theorem 
    cannot be anymore applied.  Here we can define the function
    \be
    X(C) = -{dS \over dC}, 
    \ee
    which characterizes the relations among fluctuations and response in the aging regime.
\end{itemize}

It is rather surprising to note that a dynamical version of stochastic stability \cite 
{FRMEPAPE} implies the 
relation 
\be \label{FDR}
X(C)=\int_{0}^{C} dq P(q),
\ee
which was conjectured to be valid in infinite range models \cite{CUKU1,FM}.

\begin{figure}\label{D3}
    \epsfig{figure=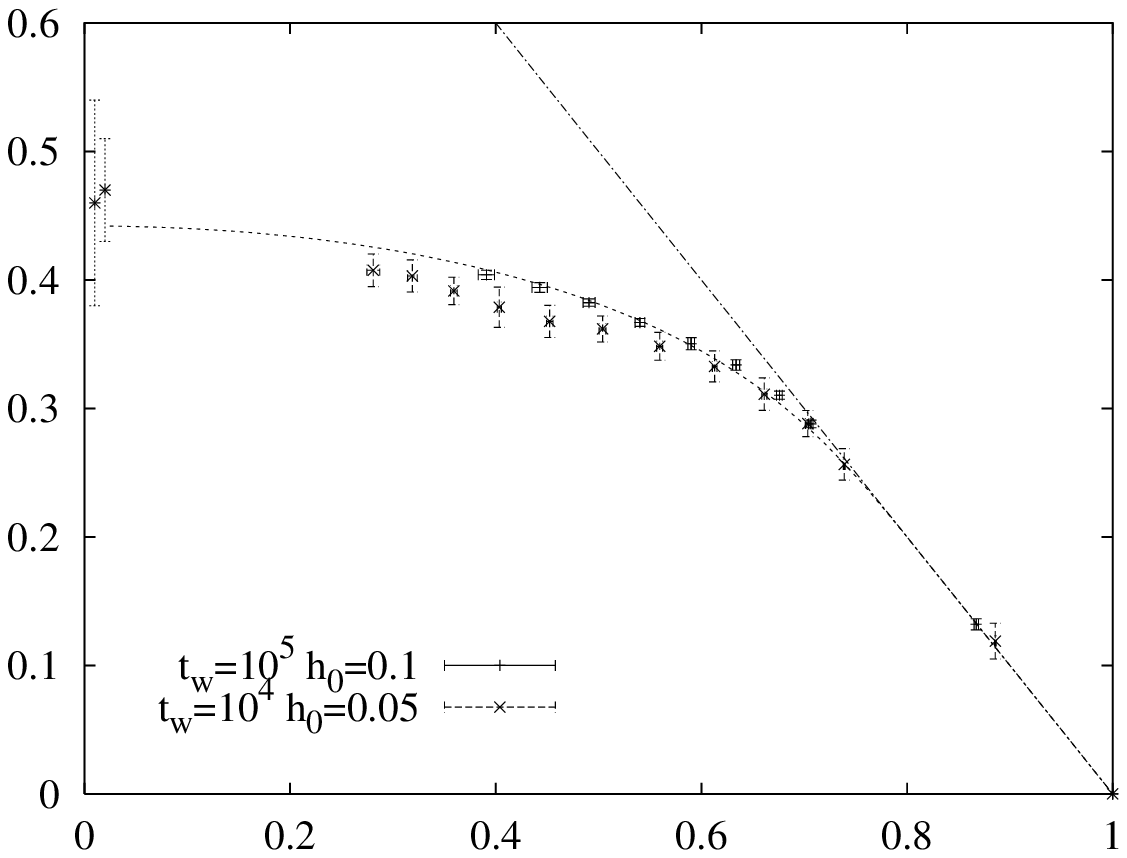,width=8cm}
    \caption{Relaxation function versus correlation in the Edwards-Anderson (EA) model in $D=3$
     $T=0.7\simeq \frac34 T_c$ and theoretical predictions from eq. (\ref{FDR}) \cite{SG}.}
\end{figure}

There are three main kinds of the dynamical response {$S(C)$}, that correspond
to different shapes of the static function {$P(q)$} as shown in fig. (\ref{TRE}).
\begin{itemize}
    \item  Case A correspond to no replica symmetry breaking and it is the usual case when two 
    phases are present (e.g.  
    ferromagnetism): $P(q)$ is a delta function.

    \item Case B correspond to one step replica symmetry breaking and it is supposed to be 
    realized in structural glasses: $P(q)$ is a the sum of two delta functions.

    \item Case C correspond to breaking the replica symmetry in an continuous way and is supposed to 
    be realized in glasses and maybe in tiling models: $P(q)$ is a the sum of two delta functions 
    plus a smooth function in between.
\end{itemize}

There are many numerical results which confirm the validity of this analysis and classification. 
Some results are presented in the next two sections.

\subsection{Numerical results for spin glasses}

\begin{figure}\label{D4}
    \epsfig{figure=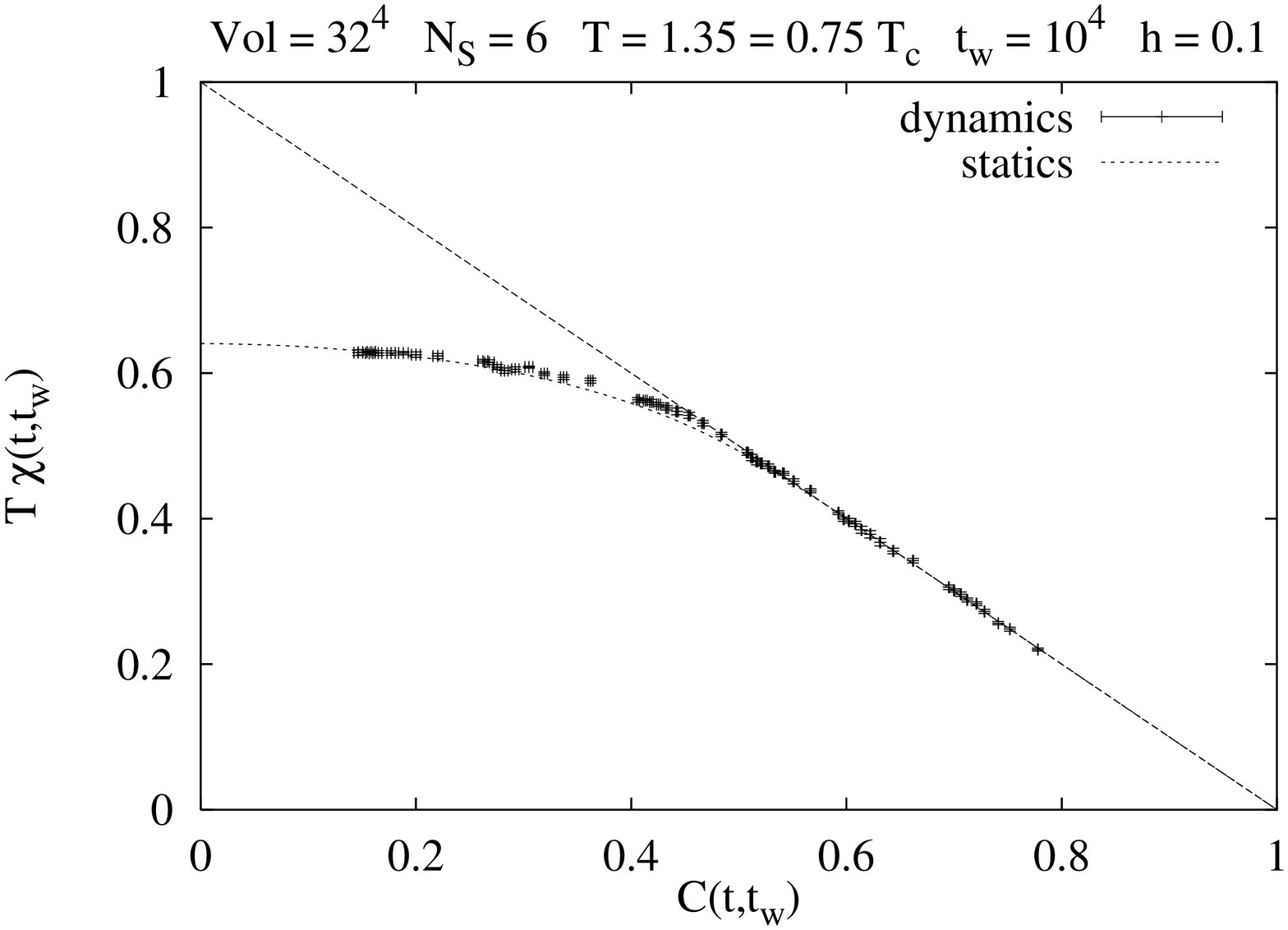,angle=270,width=8cm}
    \caption{Relaxation function versus correlation in the Edwards-Anderson (EA) model in $D=4$
     $T=0.7\simeq \frac34 T_c$ and theoretical predictions from eq. (\ref{FDR}) \cite{SG}.}
\end{figure}

In spin glasses it is possible using the parallel tempering technique to fully thermalize 
systems up to $O(10^{4})$.  System of this size could not be thermalized in a reasonable time 
(eg.  let than $10^{12}$ Monte Carlo sweep) using the conventional Monte Carlo technique.  
After thermalizing hundredths of these systems one can measure the average function $P(q)$ 
and using the relation (\ref{FDR}) one can predict the form of $X(C)$ and consequently the 
dependence of $R$ on $C$.

These predictions can be tested by taking a large sample (e.g. O($10^{6}$) spins) and by 
measuring the correlations and the relaxation function in the off-equilibrium aging 
regime.  These results can be seen in fig.  (\ref{D3}) for $D=3$ and fig.  (\ref{D4}) for 
$D=4$.  The result are quite impressive if we consider that no free parameter is present 
and that the two curves are obtained with a very different procedure.

Similar results have been obtained for a tiling model \cite{LEPA}.

\subsection {Numerical results for  structural glasses}
\begin{figure}
    \epsfig{figure=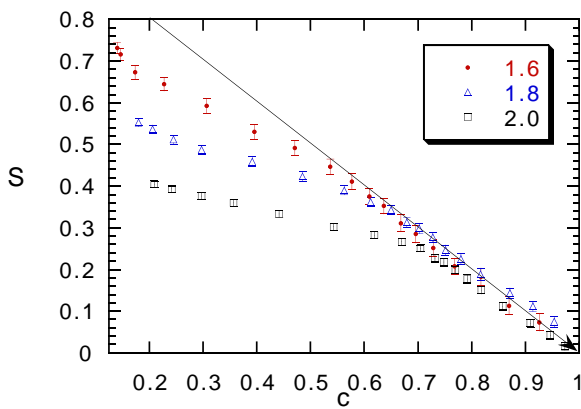,width=8cm}
    \caption{Relaxation versus correlation in a binary mixture.}
\end{figure}

In the case of structural glasses analytic computations \cite{MP} and phenomenological considerations 
\cite{KW} support the 
proposal that

\bea
X(C)=m \for C<C^{*}, \\
X(C)=1 \for C<C^{*},
\eea
where
\be
m \approx T/T_{c}
\ee
(i.e $m=1$ at $T=T_{c}$, $m=0$ at $T=0$).
The quantity
$m\beta$ is roughly independent from the temperature and has the meaning of the
effective temperature in the space of the valleys.  In other  words replica symmetry should be 
broken at one step level and the symmetry breaking parameter $m$ should be one at the critical 
temperature and vanish linearly with the temperature.

\begin{figure}\label{GLASS}
    \epsfig{figure=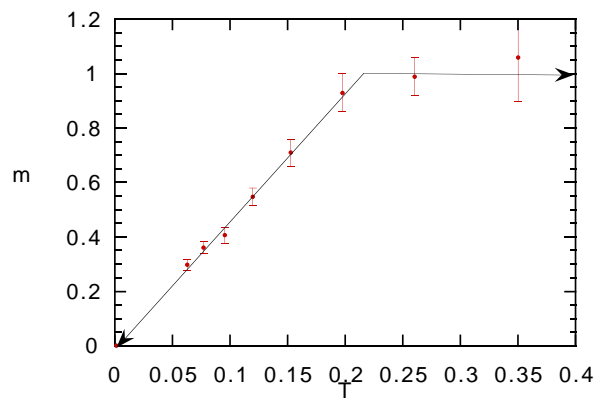,width=8cm}
    \caption{Temperature dependence of the parameter $m$ in a binary mixture.}
\end{figure}

This kind of behaviour is exactly the one that has been observed in some long range model of spin 
glasses. Numerical simulations support the correctness of these predictions. For example 
we see in fig. \ref{GLASS} the numerical evaluation of the parameter $m$ in a glass 
forming binary mixture.

\section{Conclusions}

We have argued that glassiness, metastability and the existence of many equilibrium states 
are phenomena which are strongly linked. The theory for studying these phenomena has been 
developed in the framework of replica theory, however we can abstract from the usual 
replica approach two principles, stochastic stability and separability, which are enough to 
fully characterize the theory.

The most impressive phenomenon is the presence of a new form of fluctuation dissipation 
relation in the aging regime, where the characteristic function $X(C)$ is linked to a 
quantity that can be defined in the static, i.e the function $P(q)$.  Direct measurements 
of these fluctuation dissipation relations in real (not numerical) experiments both for 
glasses and spin glasses are actually under the way.  The outcome will be a crucial test 
for this theoretical approach and I hope that some results will be available in the next 
future.

\end{document}